\documentclass{PoS}
\usepackage{amssymb, latexsym,amsmath}
\title{Neutrino propagation and quantum states in matter}
\PoS{PoS(HEP2005)191}
\author{Ivan Pivovarov,
\\Department of Quantum Field Theory and High Energy Physics,
\\Moscow State University,\\119992 Moscow, Russia
\\E-mail:\email{pivovarov\_ivan@bk.ru}}

\author{\speaker{Alexander Studenikin},
\\Department of Theoretical Physics,
\\Moscow State University,\\119992 Moscow, Russia
\\E-mail:\email{studenik@srd.sinp.msu.ru}}
\abstract{We briefly review the matter effects in case of Dirac and Majorana neutrino propagation
in medium. We develop the quantum treatment of neutrinos in matter: using the generalized Dirac
equations for Dirac and Majorana neutrinos wave functions in matter we get the explicit expressions
for the corresponding Green functions.}

\FullConference{International Europhysics Conference on High Energy Physics\\
July 21st - 27th 2005\\
Lisboa, Portugal}

\ShortTitle{Neutrino quantum states in matter}
\begin{document}
\section{Neutrino propagation in matter}
The presence of matter induce many new effects in neutrino interactions and oscillations. In
addition to the famous MSW effect \cite{wolfenstein,mikheev} of the resonance amplification of
neutrino flavor oscillations in matter, the matter can also resonantly increase neutrino spin
flavor oscillations \cite{akhmedov,marciano} in case neutrino has non-zero magnetic or transitional
moment. It was also discussed \cite{berezhiani_rossi_nunokawa} that the presence of matter can
induce neutrino-majoron interactions. In particular, neutrino decay to majoron
($\nu\to\tilde{\nu}+\phi$) and decay of majoron into the couple of neutrinos ($\phi\to 2\nu$) could
proceed in matter. Some interesting features of neutrino oscillations that are due to the motion
and polarization of background matter were also discussed in \cite{nunokawa}. In
\cite{likh_stud,egor_lob_stud} it was shown that the relativistic matter motion can significantly
shift the neutrino oscillations resonance conditions if compared with the case of non-moving
matter.

Recently within the developed quasi-classical approach to the neutrino spin evolution in the
background matter \cite{lob_stud} a new type of electromagnetic radiation produced by the Dirac
neutrino in matter was considered. We has termed this radiation the "spin light of
neutrino"($SL\nu$). In the last paper of \cite{lob_stud} $SL\nu$ was considered in gravitational
fields of rotating astrophysical objects. In these papers several astrophysical and cosmological
environments in which $SL\nu$ can give noticeable consequences were discussed. The quantum approach
to the $SL\nu$ in matter on the basis of solutions of the modified Dirac equation for a neutrino
was developed in \cite{stud_tern_grig_quant}.

The quantum treatment of Majorana neutrino moving in the background matter was discussed in
\cite{new_majorana}. Note that the neutrino wave functions and spectra in matter were also
discussed before in \cite{alltherest}.

In this paper we further develop the quantum treatment of neutrino motion in the background matter
and for the first time obtain explicit Green functions for generalized Dirac equation for the Dirac
and Majorana neutrinos.

\section{Neutrino Green function in matter}
Let us consider the electron neutrino propagating through the background matter. For moving and
unpolarized matter composed of only electrons (the electron gas) we obtain
 \cite{stud_tern_grig_quant,new_majorana} the following modified Dirac equation for the neutrino wave function:
\begin{equation}
\label{eq:GenDPeq} \bigg\{ i\gamma_\mu\partial^{\mu}-\frac{1}{2}\gamma_\mu
\left(a+\gamma_{5}\right)f^{\mu}
-m\bigg\}\Psi\left(x\right)=0,\qquad
f^{\mu}=\frac{G_{F}}{\sqrt{2}}\biggl(1+4\sin^{2}\theta_{W}\biggr)j^{\mu},\quad
j^{\mu}=\left(n,n{\bf{v}}\right).
\end{equation}
Here $n$ is the number density of the background electrons and ${\bf{v}}$ is the speed of the
reference frame in which the mean momenta of the matter electrons is zero, the parameter $a=0$ for
the case of Majorana neutrinos and $a=1$ for the Dirac neutrinos. The exact solutions of this
equation for Dirac neutrinos were found in ~\cite{stud_tern_grig_quant}, the Majorana case was
considered in \cite{new_majorana}.

The equation for the neutrino Green function is the same as Eq.(\ref{eq:GenDPeq}) with the only
difference that $-\delta(x)$ function stays on the right hand side. In the momentum representation
it has the following form:
\begin{equation}
\label{eq:greeneqp} \left(\hat p-m-\frac{1}{2}\hat f(a+\gamma_{5})\right)G(p)=-1,\quad where\quad
\hat f=f^{\mu}\gamma_{\mu},\quad\hat p=p^{\mu}\gamma_{\mu}.
\end{equation}
Squaring the left hand side operator, it is possible to obtain the following expression for the
neutrino Green function in matter:
\begin{equation}
\label{eq:vyvod_final}
G_{matt}(q)=\frac{-\left(q^{2}-m^{2}\right)\left(\hat
q+m\right)+\hat f\left(\hat q-m\right)P_{L}\left(\hat
q+m\right)-f^{2}\hat q\,P_{L}+2\left(fq\right)P_{R}\left(\hat
q+m\right)}{\left(q^{2}-m^{2}\right)^{2}-2\left(fq\right)\left(q^{2}-m^{2}\right)+f^{2}q^{2}},
\end{equation}
where
\begin{equation}
q=p-\frac{1}{2}(a-1)f,\quad q^2=q_{\mu}q^{\mu},\quad (fq)=f_{\mu}q^{\mu},\quad
P_{L}=\frac{1}{2}(1+\gamma_{5}),\quad P_{R}=\frac{1}{2}(1-\gamma_{5}).
\end{equation}

Now let us consider the denominator of Eq.(\ref{eq:vyvod_final}). The poles of the Green function
determine the neutrino dispersion relation. Equating the denominator to zero, we obtain quartic
equation relative to $q_{0}$:
\begin{equation}
\label{eq:denzero}
\left(q^{2}-m^{2}\right)^{2}-2\left(fq\right)\left(q^{2}-m^{2}\right)+f^{2}q^{2}=0.
\end{equation}
In some special cases Eq.(\ref{eq:denzero}) can be solved analytically. One of such cases is that
of uniform medium, moving at constant speed $\bf{v}$ parallel to the neutrino momentum $\bf{p}$. In
this case we can solve Eq.(\ref{eq:denzero}) for $q_{0}$, and thus for $p_{0}$ find
\begin{equation}
\label{eq:q0solution}
{p}_{0}=\frac{1}{2}\left(af_{0}+s\,|\,{\bf{f}}\,|+\epsilon\sqrt{4m^2+(2|\,{\bf{p}}-
\frac{1}{2}(a-1){\bf{f}}\,|-s(f_{0}+s|\,{\bf{f}}\,|))^{2}}\right).
\end{equation}

There are four solutions of Eq.(\ref{eq:q0solution}) with $s=\pm 1$ and $\epsilon =\pm 1$. The
value of $s$ specifies the neutrino helicity states:
\begin{equation}
\frac{\hat\Sigma {\bf{p}}}{p}\,\Psi({\bf{r}},t)=s\,\Psi({\bf{r}},t),\quad\hat\Sigma
=\begin{pmatrix}{\bf{\hat\sigma}}&0\\0&{\bf{\hat\sigma}}\end{pmatrix},
\end{equation}
whereas $\epsilon$ splits the solutions into the two branches that in the limit of the vanishing
matter density reproduce the positive and negative frequency solutions, respectively. Here
${\bf{\hat\sigma}}=(\sigma_{1},\sigma_{2},\sigma_{3})$ are the Pauli matrixes. From
Eq.(\ref{eq:q0solution}) one can find, that all solutions except one are of definite sign for any
$|{\bf{p}}|$. The sign of $p_{0}$ for $\epsilon=-1$ and $s=1$ however can be both positive and
negative for different $|{\bf{p}}|$. One can also note, that in case of
\begin{equation}
\label{eq:condition} af_{0}+|\,{\bf{f}}\,|<2m,
\end{equation}
the sign of this $p_{0}$ is always negative.
\section{Discussion}
In case condition Eq.(\ref{eq:condition}) holds, the solution of Eq.(\ref{eq:GenDPeq}) can be
expressed in the form of the superposition of plane waves each with definite sign of energy. Note
that if the condition Eq.(\ref{eq:condition}) is violated then there exists a plane wave that has
positive energy for some $|{\bf{p}}|$ and negative for others. Stated in other words,
Eq.(\ref{eq:condition}) means that Green function Eq.(\ref{eq:vyvod_final}) can be chosen causal by
imposing special rules of poles bypassing (negative poles should be bypassed from below and
positive poles should be bypassed from above). Once we got the causal Green function the
perturbation technique can be developed for the description of the neutrino propagation in matter.

Another way to interpret Eq.(\ref{eq:condition}) is to turn attention to \cite{new_majorana} where
it was shown, that for the matter at rest, the spontaneous $\nu\tilde{\nu}$ pair creation can take
place only when $f_{0}>2m$. From the analysis of the allowed energy zones for neutrino in matter it
follows that $\nu\tilde{\nu}$ pair creation in moving matter can take place only when
$af_{0}+|{\bf{f}}|>2m$. So that the possibility of using the neutrino Green function
Eq.(\ref{eq:vyvod_final}) is limited by the particular value of matter density when
$\nu\tilde{\nu}$ pair creation processes become available.

For the Majorana neutrino moving through uniform matter at rest the condition
Eq.(\ref{eq:condition}) is always valid for any matter densities $f_{0}$ because $a=0$ and
${\bf{f}}=0$ in this case.

One of the authors (A.S) is thankful to  Gaspar Barreira  and Jose Bernabeu for the invitation to
attend the International Europhysics Conference on High Energy Physics and to all organizers for
hospitality.


\begin{thebibliography}{99}
\bibitem{wolfenstein} L.~Wolfenstein, \emph{Phys.Rev.} {\bf{D17}} (1978) 2369.
\bibitem{mikheev} S.~Mikheev, A.~Smirnov, \emph{Sov.J.Nucl.Phys.} {\bf{42}} (1985) 913
\bibitem{akhmedov} E.~Akhmedov, \emph{Phys.Lett.} {\bf{B213}} (1988) 64.
\bibitem{marciano} C.~S.~Lim, W.~Marciano, \emph{Phys.Rev.} {\bf{D37}} (1988) 1368.
\bibitem{berezhiani_rossi_nunokawa} Z.~Berezhiani, M.~Vysotsky, \emph{Phys.Lett.} {\bf{B199}} (1987) 281; Z.~Berezhiani, A.~Smirnov,
\emph{Phys.Lett.} {\bf{B220}} (1989) 279; C.~Giunti, C.~W.~Kim, U~.W.~Lee, W~.P.~Lam,
\emph{Phys.Rev.} {\bf{D45}} (1992) 1557; Z.~Berezhiani, A.~Rossi, \emph{Phys.Lett.} {\bf{B336}}
(1994) 439;
\bibitem{nunokawa} H.~Nunokawa, V.~Semikoz, A.~Smirnov, J.W.F.~Walle, \emph{Nucl.Phys} {\bf{B501}} (1997)
17.
\bibitem{likh_stud} G.~Likhachev, A.~Studenikin, 1995 (unpublished);
A.~Grigoriev, A.~Lobanov, A.~Studenikin,\emph{Phys.Lett} {\bf{B535}} (2002) 187; A.~Studenikin,
\emph{Phys.Atom.Nucl.} {\bf{67}} (2004) 719 [{\tt hep-ph/0306280}]
\bibitem{egor_lob_stud} A.~Egorov, A.~Lobanov, A.~Studenikin, \emph{Phys.Lett.} {\bf{B491}},
137 (2000) A.~Lobanov, A.~Studenikin, \emph{Phys.Lett.} {\bf{B515}}, 94 (2001) M.~Dvornikov,
A.~Studenikin, \emph{JHEP} {\bf{09}} (2002) 016
\bibitem{lob_stud} A.~Lobanov, A.~Studenikin, \emph{Phys.Lett.} {\bf{B564}} (2003) 27; \emph{Phys.Lett.} {\bf{B601}} (2004)
171; M.~Dvornikov, A.~Grigoriev, A.Studenikin, \emph{Int.J.Mod.Phys.} {\bf{D14}} (2005) 308;
\bibitem{stud_tern_grig_quant} A.~Studenikin, A.~Ternov, \emph{Phys.Lett.} {\bf{B608}} (2005) 107; A.~Grigoriev, A.~Studenikin, A.~Ternov, \emph{Grav. \& Cosm.} 11 (2005) 132;
A.~Grigoriev, A.~Studenikin, A.~Ternov \emph{Phys.Lett.} {\bf{B622}} (2005) 199
\bibitem{new_majorana} A.~Grigoriev, A.~Studenikin, A.~Ternov, [{\tt hep-ph/0511330}] (to be published in \emph{Phys.Atom.Nucl}).
\bibitem {alltherest}
L.N.~Chang, R.K.~Zia \emph{Phys.Rev.} {\bf{D38}} 1669; J.~Pantaleone \emph{Phys.Lett.} {\bf{B268}}
(1991) 227; J.~Pantaleone \emph{Phys.Rev.} {\bf{D46}} (1992) 510; K.~Kiers, N.~Weiss
\emph{Phys.Rev.} {\bf{D56}} (1997) 5776; K.~Kiers M.~Tytgat \emph{Phys.Rev.} {\bf{D57}} (1998)
5970; A.~Loeb \emph{Phys.Rev.Lett.} {\bf{64}}(1990) 115; P.~Mannheim \emph{Phys.Rev.} {\bf{D37}}
(1988); A.~Kusenko, M.~Postma, \emph{Phys.Lett} {\bf{B545}} (2002) 238; M.~Kachelriess,
\emph{Phys.Lett} {\bf{B426}} (1998) 89; V.~Oraevsky, V.~Semikoz, Ya.~Smorodinsky, \emph{Phys.Lett}
{\bf{B227}} (1989) 255.
\end{thebibliography}
\end{document}